\documentclass[10pt]{article}

\usepackage{amsmath}
\usepackage{amssymb}

\usepackage{graphicx}

\usepackage{cite}

\usepackage{color} 

\usepackage[labelfont=bf,labelsep=period,justification=raggedright]{caption}

\makeatletter
\renewcommand{\@biblabel}[1]{\quad#1.}
\makeatother

\date{}

\begin{document}

\begin{flushleft}
{\Large
\textbf{Modulation of beta oscillations during movement initiation: modeling the ionic basis of a functional switch}
}
\\
Julie Dethier$^{1}$, 
Guillaume Drion$^{1}$, 
Alessio Franci$^{2}$,
Rodolphe Sepulchre$^{1,2,3,\ast}$
\\
\bf{1}   Department of Electrical Engineering and Computer Science, University of Li\`ege, Li\`ege, Belgium
\\
\bf{2}  INRIA Lille-Nord Europe, Orchestron project, Villeneuve d'Ascq, France
\\
\bf{3}  Department of Engineering, University of Cambridge, Cambridge CB2 1PZ, UK
\\
$\ast$ E-mail: jdethier@ulg.ac.be
\end{flushleft}

\section*{Abstract}
We use a computational model to propose a physiological mechanism by which transient control of beta oscillations in the indirect pathway of the basal ganglia is orchestrated at the cellular level. Our model includes a simple and robust mechanism by which a cellular switch (from bursting to tonic) almost instantaneously translates into a functional gating switch (from blocking to conducive) in an excitatory-inhibitory network. Applied to the control of beta oscillations in the basal ganglia, the model shows the modulation of beta activity under the action of a transient depolarization, for instance a dopamine signal. The model predicts, by analogy to the thalamocortical circuit, a novel gating function by which the transfer of cortical spikes through the indirect pathway is blocked under the inhibitory drive preceding movement but briefly released at the onset of movement execution.
\section*{Significance Statement}

Connecting the dots between the molecular action of a neurotransmitter and the rapid modulation of a brain network rhythm is an important challenge of neuroscience. We propose a simple, generic, and robust mechanism by which a switch mediated by neurotransmitters at the cellular level can switch off network  oscillations. We illustrate the mechanism on the modulation of beta oscillations in the basal ganglia. The model suggests a striking analogy between gating mechanisms in the basal ganglia and in the thalamocortical circuits.

\section*{Introduction}

The beta rhythm of the basal ganglia and its functional role have been the topic of increasing attention in recent years~\cite{Boraud2005, Nambu2008} primarily because of the development of deep-brain stimulation (DBS) as a treatment to alleviate the most debilitating symptoms of Parkinson's disease~\cite{Nambu2008, Hammond2007}. It is well established that the basal ganglia are composed of two parallel pathways, the direct---movements releasing---and indirect---movements inhibiting---pathways. Until two decades ago, movement disorders such as those observed in Parkinson's disease were explained by firing rate changes~\cite{Nambu2008, Albin1989}. Recently, attention has shifted to the physiological and pathological rhythmic activity in specific neuronal populations, particularly in the beta band (8 - 35Hz)~\cite{Boraud2005}. In this paper, we propose a computational model by which a cellular switch controls oscillations at the population level and modulates activity transfer gating through the indirect pathway.

The first part of the paper is purely computational and generic: we start from recent work that models a cellular switch from tonic spiking to bursting, mediated for instance by calcium channels~\cite{Franci2013a}, and explore the consequences of the cellular switch in a network composed of two interacting heterogenous populations, one excitatory and the other inhibitory. The tonic mode at the cellular level translates into an exogenous rhythm at the network level, conducive of synaptic inputs. In sharp contrast, and with no change in the synaptic connections, the bursting mode at the cellular level translates into a strongly endogenous rhythm at the network level, mostly blind to synaptic inputs. 

The second part of the paper uses a large amount of recent experimental evidence to demonstrate that the basal ganglia include all the required ingredients to provide a robust instance of the proposed mechanistic link and to support the role of T-type calcium channels as key ionic mediators between the molecular action of a transient depolarization source and the modulation of beta oscillations in the basal ganglia. At the cellular level, experimental evidence includes the prominence of T-type calcium channels both in the subthalamic nucleus (STN) and in a fraction of the globus pallidus pars externa (GPe)~\cite{Hallworth2003, Nambu1994, Cooper2000}, electrophysiological recordings of a cellular switch in neurons of the STN induced by hyperpolarization~\cite{Beurrier1999}, and the role of dopamine receptors in the striatum~\cite{Weinberger2011}. At the network level, experimental evidence includes the prominent role of the GPe-STN network in local field potential (LFP) oscillations~\cite{Plenz1999, Tachibana2011}, the modulation of beta oscillations in voluntary movements~\cite{Weinberger2011, Levy2002a, Brown2005, Sharott2005, Mallet2008a}, the link between dopamine and movements~\cite{Jin2010, Jenkinson2011}, and a number of experimentally reported effects of DBS in alleviating the abnormal oscillations~\cite{Hammond2007}.
 
We are not aware of an earlier computational model connecting the dots between the causal effects of depolarization, T-type calcium channels, cellular bursting, and beta oscillations. The model differs from previously published models of beta oscillations most radically in that a common mechanism simultaneously explains the transient physiological oscillations associated to movement control and the persistent pathological parkinsonian oscillations without any required change in the synaptic connectivity. Emphasis on the specific role of T-type calcium channels is scarce in the literature on network oscillations~\cite{Astori2011}, with the notable exception of the recent study~\cite{Tai2011} that clearly establishes their role in the basal ganglia oscillations. At a higher level, by analogy to thalamocortical circuits, the model suggests a gating of motor activity by basal ganglia circuits. The gating of sensory information has been well studied in the thalamus~\cite{McCormick1990, McCormick1997}.

\section*{Results}

\subsection*{A general mechanism for cellular control of LFP oscillations.}

\begin{figure}[t]
\centerline{\includegraphics[width=8.7cm] {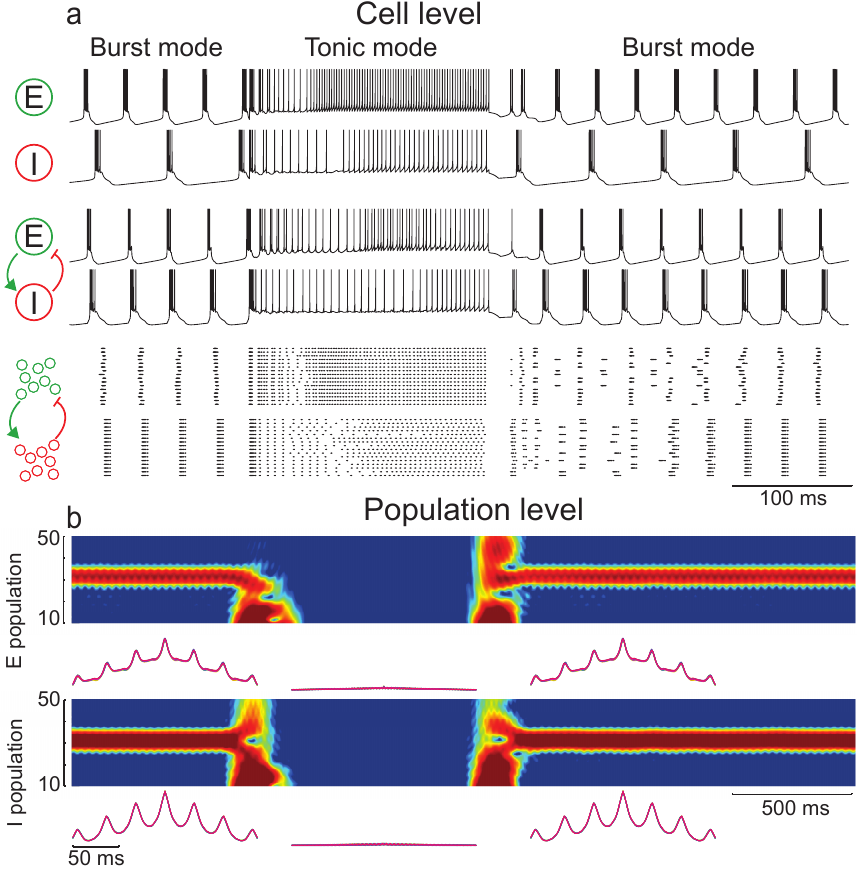}}
\caption{Regenerative excitability at the cell level robustly induces LFP oscillations at the network level. The transitions between modes are controlled by the applied current (see methods). \textbf{a}:~Cell level. Upper traces - single cells: excitability switch in two neurons with distinct intrinsic properties. Middles traces - two-neuron E-I network: in the burst mode, the two neurons synchronize and generate a network rhythm. In the tonic mode, the two neurons do not synchronize. Lower traces - E-I populations (spike rasters): the neurons fire in synchrony in the burst mode and in a decorrelated manner in the tonic mode. \textbf{b}:~Population level. First rows: LFP time-frequency plots. Oscillations at the 30 Hz frequency band develop in the burst mode but no clear oscillatory activity develops in the tonic mode. Second rows: cross-correlations for all the possible two-neuron combinations in the different modes. The burst patterns are correlated in the burst mode but not in the tonic mode. The color code (from blue to red) indicates the spectral power (from low to high).\label{fig:switch_all}}
\end{figure}

Our computational model illustrates how a cellular switch in an excitatory-inhibitory (E-I) network robustly induces LFP oscillations at a frequency that can be markedly different from the isolated cellular rhythm (Fig.~\ref{fig:switch_all}).

\textbf{Cellular switch:} The cellular switch (Fig.~\ref{fig:switch_all}a, upper traces) has been the topic of recent modeling work by the authors~\cite{Franci2013a, Drion2012, Franci2012, Franci2013}. Many neuronal models switch from tonic spiking to bursting as the balance of ionic currents in the slow time scale of repolarization evolves from negative feedback (slow restorativity) to positive feedback (slow regenerativity) at the resting potential~\cite{Franci2013}. Slow regenerative channels---most prominently, calcium channels---provide the neuron with short-term memory\cite{Marder1996a}, a key source of the hysteretic nature of bursting firing patterns~\cite{Franci2013a}. The balance can be dynamically regulated in many different ways but one prominent mechanism relevant for this paper is a transient deinactivation of T-type calcium channels by hyperpolarization~\cite{Drion2012}. A reduced low-dimensional model is sufficient to capture the cellular switch in a population computational model (see methods).

\textbf{Network switch:} In an E-I network, the cellular switch triggers a change at the population level. Fig.~\ref{fig:switch_all}a, middle traces, illustrates the rhythmic switch in a cartoon two-neuron network composed of an inhibitory neuron (I~neuron) and an excitatory neuron (E~neuron), connected reciprocally: the I~neuron is connected to the E~neuron via a GABAergic connection and the E~neuron is connected to the I~neuron with a glutamatergic connection (see methods). 

\begin{figure}[t]
\centerline{\includegraphics[width=8.7cm]{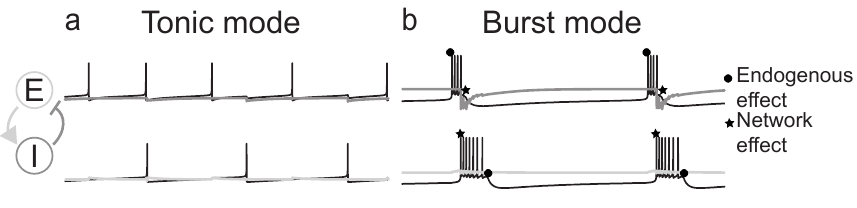}}
\caption{The network rhythm results from the combination of the cellular switch and the network coupling. The GABAergic and the glutamatergic currents are represented in dark and light grey, respectively. \textbf{a}:~In the tonic mode, there is no network rhythm because the synaptic current of a single spike is not strong enough to initiate a spike in the other neuron. \textbf{b}:~In the burst mode, the network rhythm is due to the interplay of two effects: endogenous effect due to the intrinsic properties of the neurons and network effect due to the synaptic connections.\label{fig:twoeffects}}
\end{figure}

In isolation, the two neurons have distinct intra and interburst frequencies and generate distinct rhythms (Fig.~\ref{fig:switch_all}a, upper traces). In the network configuration, the cellular switch triggers a clear switch in the network. In their tonic mode, the two neurons in the network barely affect each other because the effect of a single spike on the other neuron is weak(Fig.~\ref{fig:twoeffects}a). The two neurons, separately or connected, have almost the same firing pattern (Fig.~\ref{fig:switch_all}a, upper and middle traces, Tonic mode). The two neurons fire asynchronously. In sharp contrast, the influence of one neuron on the other becomes strong in their bursting mode. The network bursting is not a simple reflection of the unicellular bursting, but a combination of two concomitant effects: the endogenous effect and the network effect (Fig.~\ref{fig:twoeffects}b)\footnote{This is in contrast to previous work such as~\cite{Frohlich2006}.}.

As illustrated in Fig.~\ref{fig:twoeffects}, the novel network rhythm depends both on the cellular switch and on the E-I interconnection. Intrinsic properties of each neuron, not interconnection, determine the onset (through E~neuron) and the termination (through I~neuron) of the network oscillations. But the interconnection influences the duration of the network oscillation: the E burst initiates the I burst through its excitatory drive while the inhibitory drive of the I~neuron terminates the E burst earlier than in isolation. The result is that the cellular switch transmits to the network its exogenous (tonic) or endogenous (bursting) nature, but that the network interconnection regulates the frequency of the endogenous network rhythm, which can markedly differ from the frequency of the bursting neurons. It should also be emphasized that the mechanism is robust to many variants: the I burster does not need to burst in isolation but could instead be `burst excitable' (Fig.~S1b), that is, capable of bursting in response to an excitatory post-synaptic potential (EPSP) and of terminating its burst endogenously. Excitatory synapses are robust to the synapse type (see methods). Finally, the cellular control of the network switch can be through the E or I~neuron only rather than through both. In short, the network configuration does increase the robustness of the network rhythm compared to the cellular rhythm in isolation against variability of the intrinsic properties of the neuron.

\textbf{Population switch:} Fig.~\ref{fig:switch_all}a, lower traces, and Fig.~\ref{fig:switch_all}b illustrate how the network rhythm translates into LFP oscillations at the population level. Raster plots illustrate that the population switches from asynchronous behavior in tonic mode to synchronous behavior in bursting mode. The cellular switch is therefore amplified at the population level because only the endogenous bursting mode can synchronize a heterogenous population. Cross-correlograms and time-frequency plots confirm a high spectral power at 30~Hz confined to the bursting periods of the neurons.

We simulated a network with sixteen neurons in each population, random connections with gaussian white noise (each E~neuron is connected to a fixed number of random I~neurons and vice versa), and heterogenous neurons (intrinsic parameters vary in a fixed range). With identical neurons, the connection ratio between the two nuclei could be as low as $1/16$ (each neuron is connected to a single neuron in the other population which is composed of sixteen neurons) and the two populations would still synchronize and generate oscillations. With heterogenous neurons (10\% variability range in this case), the connection ratio could be as low as $1/4$ (each neuron is connected to four neurons in the other population) and the populations still synchronize. Remarkably, the parameter variability has a strong impact on the unicellular rhythm (Fig.~\ref{fig:twoeffects}b) but not on the network interactions. Therefore, a high variability in the burst onset of E~neurons is compatible with the network rhythm. The random connections, on the contrary, act on the network effects and on the burst onsets of the I~neurons and on the burst termination of the E~neurons. The cellular variability is averaged due to the presence of multiple recurrent connections. The burst duration in I~neurons is also function of the intrinsic parameters and is therefore affected by the parameters variability among neurons. Here also both the cellular switch and the network interconnection contribute to the robustness of the collective phenomenon. Many variants of the mechanism are possible depending on the network connectivity within each population. The switch control can be achieved through a fraction of the populations only provided that the heterogeneity in the cellular switch is compensated for by a sufficient level of connectivity within the population. Our experience with the computational model suggests a versatile mechanism whose robustness is only increased in larger populations.

\textbf{Functional switch:} A plausible function of the population switch is to gate activity transfer of incoming spikes through the E (or I) population~\cite{McCormick1990, McCormick1997, Sherman2001}. Fig.~\ref{fig:infotransfer}a illustrates that the endogenous (burst) rhythm of the E population carries little information about an input spike train. This is in sharp contrast with the exogenous (tonic) rhythm, where the individual pattern of action potentials is an almost faithful representation of incoming EPSPs or depolarizing inputs (Fig.~\ref{fig:infotransfer}b). The action potential pattern is function of the intensity, duration, and frequency of input spikes~\cite{McCormick1990}. In tonic firing, larger EPSPs elicit higher firing rates, in contrast to the burst mode~\cite{Sherman2001}. The tonic mode is therefore called a `transfer' (or exogenous) mode and the burst mode an `oscillatory' mode~\cite{McCormick1990}.

\subsection*{Hyperpolarization controls oscillations in the GPe-STN network.}\

\textbf{Cellular switch:} STN neurons have been shown to switch from tonic firing to bursting in isolation~\cite{Hallworth2003}. The switch is controlled by the transient deinactivation of low threshold T-type calcium currents, a slow regenerative ion channel. T-type calcium channels are inactivated in the tonic mode. However, when the neuron is hyperpolarized, the channels deinactivate and act as a strong source of regenerativity, switching the neuron to burst. Fig.~S1a reproduces, in our reduced model, a switching pattern that has been observed \textit{in vitro} (for instance, see Fig.~4~\cite{Beurrier1999}). The polarization of the membrane sets the level of inactivation of calcium channels. Some neurons in the GPe also exhibit two distinct modes of firing~\cite{Nambu1994, Cooper2000}. Although no experimental recording has shown the switch in isolation, many GPe neurons do possess the regenerative low threshold T-type calcium current~\cite{Nambu1994, Cooper2000}, and the switch can therefore be activated in models of GPe neurons as in STN neurons by modulating the balance of restorative and regenerative slow channels through hyperpolarization.

\textbf{Network switch:} Oscillations have been recorded in the GPe-STN network~\cite{Plenz1999, Tachibana2011}. The most striking evidence comes from the experimental study of~\cite{Plenz1999} showing that GPe and STN are able to generate synchronized oscillatory bursting activity \textit{in vitro}. It has also been shown that reciprocal GPe-STN interconnections, on top of glutamatergic inputs to the STN, are primordial for the generation and amplification of the oscillatory activity of STN neurons~\cite{Tachibana2011}. These considerations led to the STN/GPe pacemaker hypothesis~\cite{Bevan2002} which was implemented in numerous models~\cite{Gillies2002, Terman2002, Humphries2006, NevadoHolgado2010, Pavlides2012, Tsirogiannis2010, Kumar2011} to generate beta oscillations in the basal ganglia. The GPe-STN network fits the general mechanism of the previous section. The GPe is the I~population whereas the STN is the E~population. In this network model, the oscillatory state is controlled by the tonic input on the GPe neuron. If the GPe neuron is depolarized, both neurons are in the tonic mode and no oscillatory activity develops at the network level. By contrast, if the GPe neuron is hyperpolarized, it switches to the burst firing mode. The increased instantaneous firing rate during the burst discharge, greater than during the tonic mode (even though the global firing rate might be lower or equal to the one of the tonic mode) sets the STN neuron to an hyperpolarized level and the STN neuron crosses the excitability switch. The network oscillations start. 

\textbf{Population switch:} LFP oscillations have been recorded in the GPe-STN network~\cite{Mallet2008a, Goldberg2004}. In normal conditions, single-unit recordings display no oscillatory behavior and LFP cross-correlations show no correlated behavior, just as in the tonic mode in our model. In animal models of Parkinson's disease, the single-unit recordings reveal strong beta activity and the cross-correlations exhibit strong periodic correlated behaviors, similar to the burst mode.

\subsection*{Transient modulation of basal ganglia beta oscillations.}

The basal ganglia is a set of subcortical nuclei involved in the generation of voluntary movements. Our model targets the behavior of three of the nuclei of the indirect pathway: the striatum, the GPe, and the STN. We study the influence of dopamine modulation on the beta rhythm through the striatal D2-type dopamine receptor~\cite{Weinberger2011}.

\begin{figure*}
\begin{center}
\centerline{\includegraphics[width=17.8cm]{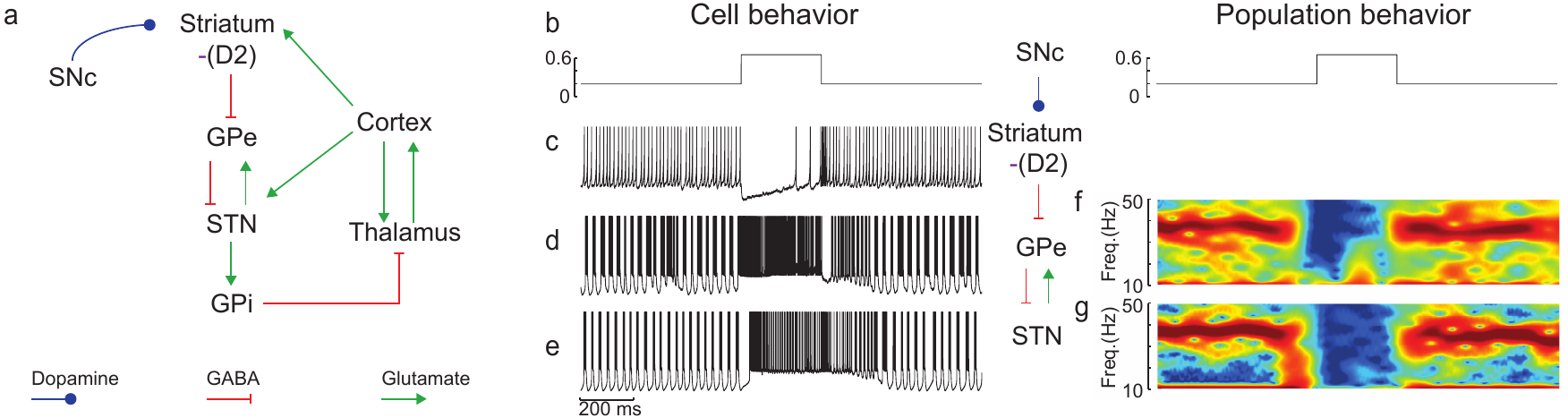}}
\caption{The indirect pathway of the basal ganglia and the dopamine control of the cellular and population behaviors in movement-related switches. \textbf{a}:~Striatal neurons with D2-type dopamine receptors, modulated by the dopamine from the SNc, make indirect contact with the output nuclei (the GPi) via relays (the GPe and STN). The output of the basal ganglia is a selectivity signal which acts on the thalamocortical activity. Blue, red, and green arrows represent dopaminergic modulatory, GABAergic inhibitory, and glutamatergic excitatory connections, respectively. SNc-substantia nigra pars compacta, GPe-globus pallidus pars externa, STN-subthalamic nucleus, GPi-globus pallidus pars interna. \textbf{b}:~Nigrostriatal kick of dopamine. \textbf{c}, \textbf{d}, and \textbf{e}:~Membrane potential of a single neuron out of the population in the striatum, GPe, and STN, respectively. The kick of dopamine decreases the firing rate of the striatal neurons which switches the GPe and the STN neurons to tonic firing. \textbf{f} and \textbf{g}:~Time-frequency plots for the GPe and STN populations, respectively. Clear beta activities exist in the GPe and STN prior to the kick of dopamine. The kick triggers an ERD before resuming to the basal activity in the beta band. The color code (from blue to red) indicates the spectral power (from low to high).\label{fig:kickDA}}
\end{center}
\end{figure*}

Our basal ganglia network simulates the dopamine modulation of the strength and coherence of the beta oscillations (see discussion). A transient increase in the nigrostriatal dopamine level from neurons in the substantia nigra pars compacta (SNc), or `kick of dopamine', triggers the decrease of the beta-band activity coherence and power~\cite{Jenkinson2011}. The unicellular switch mechanism described above is a perfect candidate due to the rapid time scales in play in movement onset (on the order of hundreds of milliseconds)~\cite{McCarthy2012, Canolty2012}. The simulated network is composed of four neurons in the striatum (only the D2-type dopamine receptors are considered) and sixteen neurons in each of the GPe and STN nucleus. The connections are as in Fig.~\ref{fig:kickDA}a. The inputs are the nigrostriatal dopaminergic level which modulates the firing rate of the striatal neurons, and cortical inputs to STN neurons. Results of the simulation show the expected behaviors (Fig.~\ref{fig:kickDA}): a kick of dopamine (Fig.~\ref{fig:kickDA}b) triggers a striatum-mediated switch in the neuronal firing rate in GPe and STN neurons (Fig.~\ref{fig:kickDA}, c, d, and e) which induces an event-related desynchronization (ERD) (Fig.~\ref{fig:kickDA}, f and g). After the movement onset, the dopamine goes back to its basal level and the striatum recovers its basal firing frequency. GPe and STN neurons undergo a reversal switch and start bursting in synchrony, generating strong network oscillations.

\subsection*{Network oscillations gate cortical activity transfer.}

\begin{figure}
\centerline{\includegraphics[width=.45\textwidth]{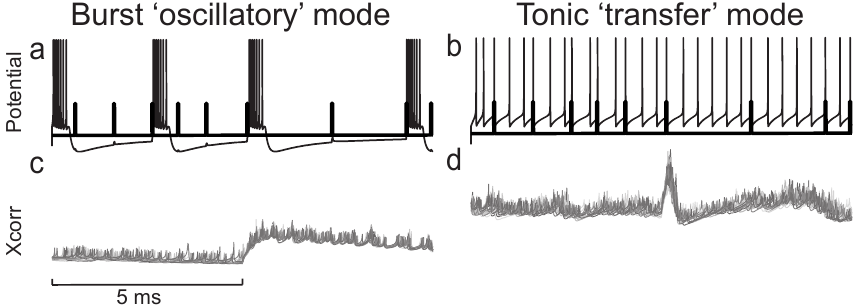}}
\caption{The two firing modes are opposite in the processing of incoming spikes. \textbf{a}~and~\textbf{b}:~A neuron membrane potential in the burst `oscillatory' filtering mode and in the tonic `transfer' mode with incoming spikes. \textbf{c}~and~\textbf{d}:~Cross-correlations (xcorr) between the membrane potential of the STN neurons and the cortical inputs. \textbf{c}:~Cross-correlations in the oscillatory network. The correlation is weak and cortical spikes are filtered. \textbf{d}:~Cross-correlations in the non-oscillatory network. The correlation is high and cortical spikes are transmitted.\label{fig:infotransfer}}
\end{figure}

The two oscillation states are opposite in their transfer of incoming spikes. Fig.~\ref{fig:infotransfer}, c and d display cross-correlations between the STN membrane potentials and cortical input spikes. The cross-correlations before the dopamine increase, when the neurons are in the burst `oscillatory' mode, possess a low maximum with no sharp transition. Input spikes from the cortex are not transmitted directly to the output firing patterns which is rather function of the internal variables. The incoming spikes might have a small impact on the interburst duration which explains the smooth increase in the cross-correlations. In sharp contrast, during the dopamine kick, when the neurons are in the tonic `transfer' mode, the cross-correlations display a sharp increase around 0 and reach a maximum of high value. Incoming spikes are transmitted to the output firing patterns, generating output spikes in STN neurons after a time delay corresponding to the synaptic transmission.

\section*{Discussion}

\subsection*{Exogenous inputs can switch the network activity from exogenous to endogenous.} 

An E-I network of neurons with sufficient expression of T-type calcium channels has the property of robustly switching from an endogenous oscillatory mode to an exogenous tonic mode under the modulation of its external drive. Hyperpolarization of the neurons deinactivates the T-type calcium channels, causing the neurons to burst and the network to oscillate, whereas depolarization inactivates the channels, causing the neurons to fire tonically and the network to stop oscillating. Both the neuron intrinsic excitability and the network interconnections participate in making the network rhythm robust and not a mere image of the isolated neuronal rhythm.

The switch is dynamic because it is controlled at the cellular level. It is robust because the E-I network oscillation is a collective synchronous phenomenon allowing for high variability at the cellular level. Because the cellular switch can be efficiently modeled in reduced neural models~\cite{Franci2013a, Drion2012, Franci2012, Franci2013}, the model is amenable to realistic simulations in large populations. The mechanism of hyperpolarization-induced bursting has been described in single-neuron electrophysiology, but we are not aware of earlier research investigating how the cellular mechanism translates to a network effect in E-I networks.

\subsection*{Beta oscillations in the GPe-STN network can be transiently switched off.}

The structure of the indirect pathway (Fig.~\ref{fig:kickDA}a) of the basal ganglia suggests that the described oscillation mechanism is consistent with the generation of beta activity in the GPe-STN network. External inputs can modulate the exogenous drive and switch the network state. For instance, an increase of cortico-striatal activity corresponds to a gain increase in the inhibitory drive from the striatum to the GPe, which might set beta activity in the GPe-STN network. However, any transient depolarization source can instantaneously switch off the beta activity. The depolarization sources are multiple, dopamine being one of them, through D2-receptors in the striatum for instance. Excitatory inputs on the STN could also be a source of depolarization.

Dopamine has been shown to correlate with the power and coherence of the beta oscillations in the basal ganglia and the cortex (in animals~\cite{Sharott2005, Mallet2008a} and in humans~\cite{Weinberger2011, Levy2002a, Brown2005}). A dopamine depletion or disrupted transmission correlates with an increase in beta oscillation power in the STN~\cite{Sharott2005, Mallet2008a}. On the opposite, dopamine therapies are associated with a decrease in the beta oscillation power and dopamine is known to suppress beta oscillations in the basal ganglia~\cite{Levy2002a, Brown2005}.

A number of experimental data are consistent with the model mechanism. At rest, the level of nigrostriatal dopamine is almost constant but differs from zero (Fig.~2e of~\cite{Jin2010}). An increase in cortico-striatal activity can set the GPe-STN network in its oscillatory state. When a kick of dopamine is produced by nigrostriatal dopaminergic neurons (Fig.~2e of~\cite{Jin2010}), the firing rate of the striatal neurons decrease drastically (Fig.~2b of~\cite{Jin2010}). The tonic inhibition on the GPe neurons is decreased and allows the neurons to switch from their burst mode to their tonic firing mode. The same phenomenon takes place in the STN neurons. GPe and STN neurons fire asynchronously and no global oscillatory activity is detected at the network level~\cite{Weinberger2011, Levy2002a, Brown2005, Engel2010, Kuhn2006, Kuhn2004}. After a few hundreds of milliseconds, the kick of dopamine ends and the GPe and STN neurons resume to their initial bursting state.

\subsection*{Relationship to previous basal ganglia models.}

The combination of a unicellular switch in individual neurons and its emulation at the population level in an E-I network is a central and novel feature of the proposed model, which, to our knowledge, is the first to investigate the dynamic modulation of beta oscillations in physiological conditions. Several earlier models have investigated GPe-STN oscillations in the pathological framework of Parkinson's disease~\cite{Gillies2002, NevadoHolgado2010, Pavlides2012, Tsirogiannis2010, Terman2002, Humphries2006, Kumar2011}. In firing-rate models~\cite{Gillies2002, NevadoHolgado2010, Pavlides2012, Tsirogiannis2010}, the modulation of oscillatory activity in an E-I network involves changes in the connectivity strength. Such changes cannot be the result of synaptic plasticity in the rapid modulation of beta activity but they could be regarded as a mean-field abstraction of the cellular modulation described here. Spiking neural models have emerged in the recent years to model network oscillatory activity~\cite{Terman2002, Humphries2006, Kumar2011}. In these models, individual neurons are usually of the leaky integrate-and-fire type, which prevents to capture an excitability switch at the neuronal level. As a consequence, network oscillations in these models are also modulated indirectly via a change of connectivity. The closest such effort to our model is described in~\cite{Kumar2011}, where modulation of dopamine levels triggers network oscillation in a GPe-STN network.

\subsection*{Transient beta oscillations and relation to movements.}

The basal ganglia play a major role in the planning and initiation of voluntary movements. In particular, the coherence and strength of the beta oscillations in the basal ganglia are associated with voluntary movements initiation, both in animals~\cite{Engel2010, Leventhal2012} and humans~\cite{Levy2002a, Brown2005, Engel2010, Kuhn2006, Amirnovin2004, Kuhn2004}. Experimental studies show a reduction or abolishment of oscillatory activity at the single-cell level during movements~\cite{Amirnovin2004}. At the population level, an ERD of oscillatory beta activity in the STN region is observed at the movement initiation~\cite{Levy2002a, Kuhn2006}. Additionally, the ERD onset latency strongly correlates with the mean reaction time in reaction-time tasks~\cite{Kuhn2004}. After the movement onset, an event-related synchronization (ERS) is observed in the beta band~\cite{Kuhn2006}.

Initiation of voluntary movements is also linked to an increase in dopamine and, in particular, to a transient increase in the activity of nigrostriatal circuits in response to cues in reaction-time tasks or prior to voluntary movements~\cite{Nambu2008, Korchounov2010, Jin2010, Jenkinson2011}. The increase in the firing rate of SNc dopaminergic neurons lasts on the order of hundreds of milliseconds and happens prior to the movement execution (Fig.~2e of~\cite{Jin2010},~\cite{Jenkinson2011}). Striatal neurons with D2-type dopamine receptors follow the inverse trend and decrease their firing rate for a few hundreds of milliseconds, covering the movement duration (Fig.~2b of~\cite{Jin2010}).

\subsection*{Pathological oscillations.}

The hallmark of Parkinson's disease is a dopaminergic denervation of the striatum, input stage of the basal ganglia, altering information patterns along movement-related ganglia-mediated pathways in the brain. Severe motor symptoms result~\cite{Hammond2007, Weinberger2009}. In Parkinson's disease, the baseline level of synchrony within and between basal ganglia nuclei in the beta band is elevated: increased beta-band activity has been reported in the STN, GPe, and GPi nuclei in both single-unit activity and LFPs~\cite{Boraud2005}. Similarly, animal models of the disease have shown increases in basal ganglia structures in burst discharges, oscillatory firing, and synchronous firing patterns~\cite{Rubin2012}. The excessive synchrony in the beta frequency band correlates with the motor symptoms and is believed to limit the information coding of neurons, the neurons being locked to the population rhythm~\cite{Hammond2007, Brown2005}. Beta synchrony diminution under dopaminergic therapies, ablative surgeries, or during DBS is associated with an amelioration of the motor impairments~\cite{Hammond2007, Weinberger2009}. It has been suggested that the degree of suppression of beta oscillations in the STN by dopaminergic medications can predict the level of improvement in bradykinesia and rigidity~\cite{Weinberger2009}.

Parkinson's disease corresponds to a pathological depleted dopaminergic state. Normally, at the onset and during movement, the synchronization in basal ganglia structures is drastically reduced. But in parkinsonian conditions, baseline levels of synchrony are higher and more resistant to suppression~\cite{Hammond2007}. In this configuration, the transient depolarization from the dopamine kick is not sufficient to trigger a network switch and strong beta oscillations remain~\cite{Jenkinson2011}, impairing cortical spikes transfer through the basal ganglia and motor movements. Our model is fully consistent with these observations and offers further insights: a decrease in the tonic nigrostriatal dopamine level projecting on neurons in the striatum results in a higher striatal firing rate. GPe neurons have a higher tonic inhibition and are maintained in the burst mode. The same observation holds for neurons in the STN. A kick of dopamine is not sufficient to bring GPe neurons to a depolarized state and to a tonic firing mode. The burst filtering mode is maintained and cortical activity cannot be processed faithfully through the basal ganglia. Prolonged synchronization among neurons in the basal ganglia, particularly between the GPe and STN, can generate a change in connection strength via synaptic plasticity, stabilizing the pathological dynamics in the network, which coincides with experimental observations~\cite{Albin1989}.
 
Experimental evidences show that T-type calcium channels, as employed in our model, are strongly involved in the disease. Indeed, pharmacological blockade of T-type calcium channels decreases burst activity in STN neurons both \textit{in vitro} and \textit{in vivo} and reduces motor deficits in animal models~\cite{Tai2011, Xiang2011}. In addition, low-frequency DBS, with long depolarization pulses, improve motor impairments in parkinsonian rats presumably by inactivating T-type channels~\cite{Tai2011}. In our model, this corresponds to bringing STN neurons to a depolarized level and tonic firing mode.

\subsection*{Analogy to thalamocortical oscillations and gating functions.}

The oscillation mechanism proposed in the present paper can also be relevant to network oscillations in the thalamocortical circuit, in particular in the thalamocortical relay-thalamic reticular network. At the individual level, the cellular switch has been described in both types of cell and is also controlled through deinactivation of T-type calcium channels~\cite{McCormick1997}. At the network level, the two populations switch from a tonic exogenous mode to an oscillatory endogenous mode and the resulting network oscillations have been proposed as the neural correlates of sleep-spindles~\cite{McCormick1997}. In contrast to our approach, however, existing computational models of the thalamocortical network~\cite{Destexhe1993a, Timofeev2005, Wang1995} do not provide a mechanism to explain the population switch. Instead, they focus on the oscillation mechanism solely in the oscillating state. In thalamic cells, the activation loss comes from a series of neuromodulators (acetylcholine, norepinephrine, and serotonin) involved in slow-wave sleep and anesthesia~\cite{McCormick1997, Llinas2006}. 

The gating function of those thalamocortical network oscillations has been studied throughly~\cite{McCormick1990, McCormick1997}. The analogy suggests a similar gating function in the basal ganglia, consistent with the modulation of beta activity during movement preparation and execution. The oscillatory network state functions as an open or closed gate towards the transfer of incoming cortical inputs through the indirect pathway. The two opposite activity processing states give a functional significance to the ERD in the basal ganglia prior to voluntary movements. During movement preparation, the power of the beta band is high and the basal ganglia-thalamocortical circuit is unresponsive to motor stimuli. The indirect pathway provides a strong inhibitory drive to the thalamus as a result of the inhibitory synchronized bursts and cortical spikes are not transferred through the indirect pathway. Right before movement onset, the sharp decrease in the beta band sets the basal ganglia-thalamocortical circuit in a responsive state, both suppressing the inhibitory drive of the indirect pathway and allowing for faithful transfer of cortical spikes. The potential functional role of beta-band oscillations could therefore be to maintain the current sensorimotor state during movement preparation~\cite{Engel2010}, until dopamine signals the onset of movement execution~\cite{Jenkinson2011}. It will be of interest to further investigate the generality of the proposed mechanism in gating population switches.



\section*{Materials}
All the numerical simulations and analyses were performed with MATLAB, MathWorks. More details on the model and analyses are given in the \textit{SI}.

\subsection*{Computational model} The neuron model used in this paper is a modified version of the model described previously~\cite{Franci2013a, Drion2012, Franci2012, Franci2013}. The effect of T-type calcium channels on the neuron excitability is accounted for by a voltage dependence of the bifurcation parameter, $w_0$, which reflects the ultraslow deinactivation of T-type calcium channels at low threshold (see \textit{SI}). A gaussian white noise is added to the system (see \textit{SI}). The synaptic connections are made with exponential synapses, the GABAergic connections being of the GABA$_{\textrm{A}}$ type, the glutamatergic of the AMPA type (see \textit{SI}) for the E-I network, and of the NMDA type for the basal ganglia network(see \textit{SI}). The generic E-I network shows robustness toward the glutamatergic synapse type (AMPA or NMDA) provided that the excitatory current ends before the burst onset of the I neuron.
The variability of the intrinsic parameters concerns the slow time-scale, $\epsilon$, the ultraslow time-scale, $\epsilon_z$, and the ultraslow equivalent gain, $k_z$. Parameters were drawn from a uniform distribution covering a 10\% range around the initial parameter value. The robustness of the switch control was tested by varying the value of $w_{0,min}$.
In the generic E-I network (Fig.~\ref{fig:switch_all}), the cellular and network switches are controlled by the applied current, $I_{app}$, in both populations, which controls the value of $w_0$ by depolarizing or hyperpolarizing the neuron. 
In the basal ganglia network, the GPe-STN connections are made in an all-to-all manner. The synaptic current, $I_{syn}$, results from the synaptic connections and controls the value of $w_0$. The striatal neuron model includes the effect of dopamine on D2-receptors via a scaling coefficient for the striatal applied current (see \textit{SI}). Striatal neurons connect to GPe neurons in an all-to-all manner with GABA$_{\textrm{A}}$ synapses. Cortical inputs on the STN neurons are spikes with spike intervals selected from a Poisson distribution, connected with AMPA synaptic connections (see \textit{SI}). We emphasize that our objective in the present paper is not a fine tuned quantitative modeling of the neuron firing pattern. Rather, we attempt to provide a qualitative picture of the switch behavior of neurons and how this switch behavior impacts neural activity processing properties. For this reason, the firing rates, burst durations, and interburst durations are not fitted to the ones observed experimentally.

\subsection*{Analyses} LFPs dynamics are modeled by the low-pass filtered (<100Hz) normalized sum of the collective synaptic activities (see \textit{SI}). The time-frequency plots result from the logarithmic representation of the LFP spectrogram obtained via a short-time Fourier transform. The cross-correlations are obtained from the MATLAB cross-correlation function estimate, for all the combination of two neurons' membrane potential (Fig.~\ref{fig:switch_all}) or the input cortical spike train and all the STN neurons' membrane potential (Fig.~\ref{fig:infotransfer}).


\section*{Acknowledgments}
The authors would like to thank T. Boraud, P. Maquet, and V. Seutin for their helpful comments. This paper presents research results of the Belgian Network DYSCO (Dynamical Systems, Control, and Optimization), funded by the Interuniversity Attraction Poles Programme, initiated by the Belgian State, Science Policy Office. The scientific responsibility rests with its authors. J. Dethier is supported by the F.R.S.-FNRS (Belgian Fund for Scientific Research).

\bibliography{pnas_bilbi}

\end{document}